\documentclass{amsart}
\usepackage{graphicx}
\usepackage{epstopdf}
\theoremstyle{definition}

\theoremstyle{remark}

\numberwithin{equation}{section}

\begin{document}
\title[]{Remarks on the statistical aspects of the safety analysis}
\author{L\'en\'ard P\'al$^{1}$ and Mih\'aly Makai$^{2}$}
\address{$^{1}$Energy Research Center, Hungarian Academy of Sciences, H-1525 Budapest 114, POB 49,
Hungary\\ $^{2}$ BME Institute of Nuclear Techniques, H-1111 Budapest, Muegyetem rkp. 3-9}
\email{$^{1}$lpal\@rmki.kfki.hu $^{2}$makai\@reak.bme.hu}

\thanks{}%

\date{\today}%
\begin{abstract}
We investigate the statistical methods applied
throughout safety analysis of complex systems. The tolerance
interval method implemented in the widely utilized 0.95|0.95
methodology is analyzed. We point out a remarkable weakness of the
tolerance interval method concerning the principle of repeatability.
It is proved that repeating twice the procedure, the probability
that the second maximum  will be larger/smaller than the first one
is 50\%. This statement is not surprising, it holds for any random
variable with continuous distribution function. In order to
demonstrate the undesirable consequences of the tolerance interval
method in the decision making, the results of the analysis of an
elementary example are discussed. Instead of the tolerance interval
method, we suggest another method based on the sign test which has
more encouraging features, especially in the case of several output
variables. The problematic aspects of the method are also discussed.
Finally, we suggest a simple test case which is able to reveal if
the tolerance interval method would not be capable of determining the
risky states of the system, when there are only a few of them. If
there are many, then the method may not explore each one in the
analysis.
\end{abstract}
\maketitle
\section{Introduction}

Regulation of industrial devices prescribes an analysis of the risk
associated with the operation of the device. The US regulation
(10CFR §50.46), which is considered as standard world wide, fixes
the requirement that the analyst should conclude "with high
probability" that the device is safe. At the same time, one can add
further self-explaining requirements, like
\begin{itemize}
    \item \textit{Repeatability}: when the analysis is repeated one should get
    more or less the same safety limits or other relevant safety features;
    \item \textit{Objectivity}: the analysis should be based on scientific
    considerations, not on individual considerations.
    This assumes that principles of safety analysis are widely discussed
    and approved by experts.
    \item \textit{Transparency of the procedure}: The safety analysis is prepared
    by a group of experts, and is discussed by experts of several professions.
    It should be assured that there are participants in the discussion
    who are capable of providing pieces of information on every relevant
    and major areas of the safety analysis (plant data, design principles,
    and methodology of the analysis).
\end{itemize}
The first two items touch the safety analysis itself, whereas the
third one touches rather organizational problems. The present work
deals with the first two items. The analyst \cite{kry90}  may
utilize the best estimate method with uncertainty analysis. Here a
possible approach is to select a device (e.g. a nuclear reactor, chemical reactor)
with nominal parameters and to specify those parameters which are
involved in the uncertainty analysis. In the course of the analysis, the analyst selects a possible realization of the device by replacing the nominal parameters by actual parameters and calculates the parameters to be subjected to limitations. We call this procedure a \emph{run}. The actual parameters are drawn from a probability distribution determined by engineering judgement.

We are going to point out that the traditional $0.95 \vert 0.95$ tolerance
interval method fails to be repeatable and objective because the
repeated code runs may lead to considerable differences in the obtained
maxima. In a considerable portion of the cases repeating for instance twice the run, the probability that
the maximum obtained in the second run will be larger/smaller than
in the first run is $1/2$. This statement is not at all surprising,
since it holds for any random variable with continuous distribution
function. \textbf{The problem is that the  random nature of the maximal
value is often disregarded in practice.}

We point out the weakness of the $0.95 \vert 0.95$ methodology
by presenting a simple example in which the "code for simulation"
is replaced by random outputs obeying lognormal distribution,
and analyze the properties of samples taken from these outputs.

Instead of the $0.95 \vert 0.95$ methodology we propose another
statistical method, called sign test, which is a variant of the good old
Clopper-Pearson \cite{clpr34} method. We claim that the method proposed by us has some
advantages in comparison with the tolerance interval method, furthermore the
erratic behavior of the coverage probability of the confidence
interval \cite{brown01} renders the method conservatism and puts the
analyst on the safe side. The main advantage of the sign test method is
nothing else than it is directly applicable in the case of several
output variables.

Finally, we analyze a trivial test problem in the usual safety
analysis frame: a series of calculations are performed by a
numerical model of the device. The analysis ought to verify that no
limit violation occurs in any realistic device state. In the
suggested test, limit violation occurs  only in 1\% of the possible
states. The challenge is to pinpoint the risky states.

The structure of the present work is as follows. In Section 2, we
analyze the random nature of the maximal values of the sample set produced
by runs of a code as it is usually done with any best estimate method. In Section 3, we present an
elementary example to point out the drawbacks of this method. The
sign test is discussed in Section 4, and in Section 5, we recommend a
simple test to verify the statistical methods to be applied in statistical inference used in safety
analysis. Conclusions are given in Section 6.

\section{Features of the tolerance interval method}

We are going to analyze the safety of a device. It is assumed that there is a computer model associated with the device. The computer model is assumed to be a best estimate model\footnote{The criteria of the decision about the acceptance
of a computer code as the best estimate method are extremely
important, but in the present work we do not deal with them.}. In the
first step the major uncertain input parameters should be
identified, their probability density functions have to be
determined, usually by engineering judgement. The model provides us with output variables which are random variables because of the random input these are subject to limitations. The joint probability distribution
function of the output is unknown. By repeatedly running the code, the analyst obtains a data set representing the operation of the device which is regarded now a "black box" device.
By exploiting the information in the data set, the analyst
tries to make a decision whether the device is safe or not.
Generally accepted, that the tolerance interval method, known simply
under the name $0.95 \vert 0.95$ methodology, can be used for this
purposes. In former papers \cite{gpm03}, \cite{pm06} we discussed
the tolerance interval method in detail. In the present article, we will use only a simple version of the mentioned method  to demonstrate  its weakness  arising because  of a random parameter  involved in the safety criteria. Unfortunately most analysts neglect  aftermath of that fact.

Before discussing the statistical aspects of the problem, we sum up the  analyst's riddle. With the code at his disposal, the analyst determines the parameters under limitation. The key question is: whether the calculated parameters exceed the limits or not. In the analysis a major problem is that the device is not exactly determined, the device parameters lie in an interval. This is taken into account by selecting randomly the actual state of the device. The question is:  had we repeat the calculations with different parameters, would we have get a more risky state or not? The resolution of the analyst problem is based on statistical considerations and is given below.

Let $y$ be a single random output, which is subject to limitation.
Let the acceptance range be given as $(-\infty, U_{T}]$, where
$U_{T}$ is the technological limit for $y$. We assume that the
distribution of $y$ is unknown, and are looking for a quantile
$Q_{\gamma}$ such that
\begin{equation} \label{1}
\int_{-\infty}^{Q_{\gamma}} dG(y) = \gamma,
\end{equation}
where $G(y)$ is the cumulative distribution function of the output
variable $y$. Quantile $Q_{\gamma}$ is to be estimated from calculations, thus, itself is a random variable. Let us consider the
results of $N$ runs of a code modeling the output variable $y$. The
$N$ values obtained in $N$ runs form a sample, and let us produce
$n+1$ samples. The first sample $ y_{01}, y_{02}, \cdots, y_{0N}$
will be called basic sample. Introducing the ordered samples
$$ y_{k}(1) < y_{k}(2) < \cdots < y_{k}(N),\;\;\;\;\;\; \mbox{where}
\;\;\;\;\;\; k = 0, 1, \ldots, n,$$ we can write the sample elements
into the following $(n+1)\times N$ matrix:
\begin{equation}\label{2}
\begin{array}{cccc}
      y_{0}(1) & y_{0}(2) & \cdots & y_{0}(N) \\
      y_{1}(1) & y_{1}(2) & \cdots & y_{1}(N) \\
      \cdots & \cdots & \cdots & \cdots \\
      y_{n}(1) & y_{n}(2) & \cdots & y_{n}(N),
\end{array}
\end{equation}
in which we call the first row \emph{the ordered basic sample}. Assuming that
the unknown cumulative distribution function $G(y)$ is monotonously
increasing and continuous, it can be immediately proved the well
known statement that
\begin{equation} \label{3}
\mathcal{P}\left\{y_{k}(N) > G^{-1}(\gamma) \right\}=
\mathcal{P}\left\{\int_{-\infty} ^{y_{k}(N)}dG(y) > \gamma \right\}
= 1 - \gamma^N = \beta,
\end{equation}
where $G^{-1}(\gamma)=Q_\gamma$ is the $\gamma$-quantile of the
probability distribution function $G(y)$. In other words,  any upper
sided interval $[-\infty, y_{k}(N)]$ covers more than the
$\gamma$-quantile $Q_\gamma$ of the output variable $y$ with
probability $\beta$.~\footnote{It is worth mentioning that the probability
of covering  more than the $\gamma$-quantile of the distribution
function $G(y)$ by any upper sided interval $[-\infty, y_{k}(N-s)]$
is given by the equation $$ \beta = I_{1-\gamma}(N-s+1, s),$$ where
$I_{1-\gamma}(N-s+1, s)$ is the cumulative beta-distribution
function.}

\textbf{Since one finds misinterpretations in the engineering practice it is
not superfluous to underline once again the proven notion of formula}
(\ref{3}). $\beta $ is the probability that the largest value
$y_{k}(N)$ of the $k$th  sample comprising $N$ output values is
greater then the $\gamma$ quantile of the unknown distribution
$G(y)$ of the output variable $y$. Another formulation asserts that
$\beta $ is the probability that the interval $[-\infty, y_{k}(N)]$
covers a larger than $\gamma$ portion of the unknown probability
distribution function. The $\gamma$ is often called as probability
content and $\beta$ as confidence level.

The maximal values $y_{0}(N), y_{1}(N), \ldots, y_{n}(N)$ are
independent and identically distributed nonnegative continuous
random variables. Hence, if $k \not= \ell $, where $k \leq n$ and
$\ell \leq n$ are nonnegative integers, then the probability of the
event $\left\{y_{k}(N) < y_{\ell}(N)\right\}$ is equal to that of
the event $\left\{y_{k}(N) > y_{\ell}(N)\right\}$, and one obtains
immediately that
\begin{equation} \label{4}
{\mathcal P}\left\{y_{k}(N) < y_{\ell}(N)\right\} = {\mathcal
P}\left\{y_{k}(N) > y_{\ell}(N)\right\} = \frac{1}{2}.
\end{equation}
Introduce the notation
\begin{equation} \label{5}
{\mathcal P}\left\{y_{k}(N) < y \right\} = {\mathcal
P}\left\{\max_{1\leq j \leq N} y_{kj} \leq y\right\} =
\prod_{j=1}^{N} {\mathcal P}\left\{y_{kj} \leq y \right\} =
\left[G(y)\right]^{N} = H(y),
\end{equation}
$$ k = 0, 1, \ldots, n, $$
one can write that
$$ {\mathcal P}\left\{y_{k}(N) < y_{\ell}(N)\right\} =
\int_{-\infty}^{+\infty} \left[1 - H(y)\right]\;dH(y) = \frac{1}{2}
$$ and similarly $$ {\mathcal P}\left\{y_{k}(N) >
y_{\ell}(N)\right\} = \int_{-\infty}^{+\infty}H(y)\;dH(y) =
\frac{1}{2}.
$$
From the equation (\ref{4}), the trivial statement follows that the probability of
finding the maximal value in a sample larger/smaller than in the
previous one, is $1/2$.

Let $0 \leq \nu_{n}(y) \leq n$ denote the number of those random
variables from among $$y_{1}(N), \ldots, y_{n}(N)$$ which are
greater than $y$. Obviously,
\begin{equation}\label{6}
\mathcal{P}\left\{\nu_{n}(y)=\ell \right\} = J_{\ell} ^{(n)}(y) =
\binom{n}{\ell}\;\left[ 1-H(y)\right]^{\ell} \left[
H(y)\right]^{(n-\ell)}.
\end{equation}
Let $p_{\ell}^{(n)}$ stand for the probability that from among the
random variables $y_{1}(N), \ldots, y_{n}(N) $ exactly  $\ell \leq
n$ is greater than $y_{0}(N)$, which may be any real number. We
obtain from (\ref{6}) that
\begin{equation} \label{7}
p_{\ell}^{(n)} = \int_{-\infty}^{+\infty} J_{\ell} ^{(n)}(y\;dH(y) =
\binom{n}{\ell} \;\int_{0}^{1} (1-u)^{\ell}\, u^{n-\ell}\;du =
\frac{1}{n+1}.
\end{equation}
Note, that the probability $p_{\ell}^{(n)}$ is independent of
$\ell$.~\footnote{This statement is valid for any random variable of
continuous distribution.} The probability $P_{k}^{(n)}$, that at
least $k \leq n$ out of maximal values $y_{1}(N), y_{2}(N), \ldots,
y_{n}(N)$ will exceed the basic maximal value $y_{0}(N)$, can be
calculated by using the formula (\ref{7}). We have that
\begin{equation} \label{8}
P_{k}^{(n)} = \sum_{\ell=k}^{n} p_{\ell}^{(n)} = \frac{n+1-k}{n+1} =
1 - \frac{k}{n+1},
\end{equation}
which is independent of $\gamma$ and $\beta$. In the sequel the
larger/smaller statements will be used only in their "larger"
interpretation.

Let $\lambda$ be the number of those $y_{1}(N), y_{2}(N), \ldots,
y_{n}(N)$ variables which are larger than the basic $y_{0}(N)$.
Clearly, $\lambda$ is a random variable, its expectation value is
\begin{equation} \label{9}
\mathbf{E}\{\lambda\}= \sum_{\ell=1}^{n} \ell\; p_{\ell}^{(n)} =
\frac{n}{2},
\end{equation}
and its variance is $$ \mathbf{D}^2\{\lambda\}= \sum_{\ell=1}^{n}
\left(\ell-n/2\right)^{2}\; p_\ell =
\frac{1}{6}\,\left(1+\frac{n}{2}\right).$$ As a consequence, we see
that the expectation of the number of maximal values out of $n$
samples, that will exceed the basic maximal value $y_{0}(N)$, is
50\% of $n$, i.e. repeating the run many times we may expect maximal
values higher than the basic maximal value in every second case.

In the light of this statement one asks: is this the intended
outcome of the $0.95 \vert 0.95$ methodology? A further serious
objection to the $0.95 \vert 0.95$ methodology is the absence of the
uniqueness.
\begin{figure} [ht!]
\protect \centering{
\includegraphics[scale=0.9]{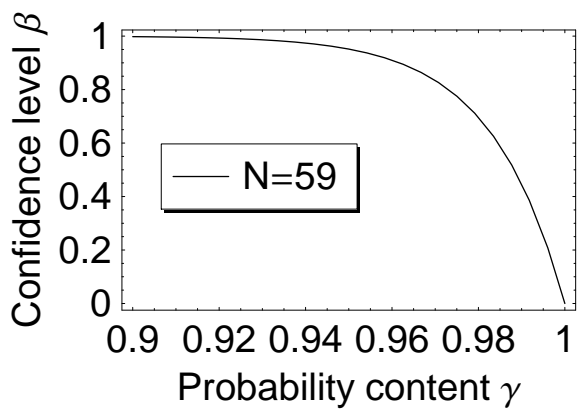}}\protect
\vskip 0.2cm \protect \caption{\label{figx}{\footnotesize Dependence
of the confidence level $\beta$ on the probability content $\gamma$
for sample size $N=59$.}}
\end{figure}
In Fig. \ref{figx} on can see the dependence of the confidence level
$\beta$ on the probability content $\gamma$ for sample size $N=59$.
The statement that the maximal element in a sample obtained by
$N=59$ runs is higher than the unknown quantile $Q_{0.95}$  with
probability $0.95$ is equivalent, for instance, to the statement
that the maximal element is higher than the unknown quantile
$Q_{0.98}$ but with probability $0.7$. It is clear that there are
infinite many $\beta \vert \gamma$ pairs corresponding to the same
$N$ value. In the safety analysis the $0.95 \vert 0.95$ pair is
accepted  but not mentioned that other pairs are also allowed.

\section{A witness of the weakness}

\textbf{In order to demonstrate the weakness of the tolerance interval
method}
, we consider a rather simple example. Take a single
output variable $y$ of lognormal distribution~\footnote{The
type of the distribution function does not have any substantial
influence on the considerations.} with parameters $m$ and $d$. This
plays the role of our "unknown" $G(y)$ distribution. The density function is
\begin{equation} \label{10}
g(y) = \frac{1}{yd \sqrt {2\pi}} \exp \left[- \frac{1}{2} \left(
\frac{\log y - m}{d}\right)^{2} \right],
\end{equation}
where $y \geq 0$.

We carried out the following numerical experiment. By means of Monte Carlo simulation we generated samples of size $N=59$. In the
simulation we have taken $m=2.5$ and $d=0.5$.  As mentioned before,
we call the first sample generated by $N=59$ runs the basic
sample, the maximal element of which is denoted  by $y_{0}(59)$.
Then, we repeat the sample generation $n=1000$ times, and select the
maximal value out of each sample. We obtained a series
$$y_{1}(59), \ldots, y_{1000}(59)$$ of the upper limits of the one-sided
tolerance intervals which cover the $95$\% of the distribution with
probability $\beta = 0.95$.

\begin{figure} [ht!]
\protect \centering{
\includegraphics[scale=0.9]{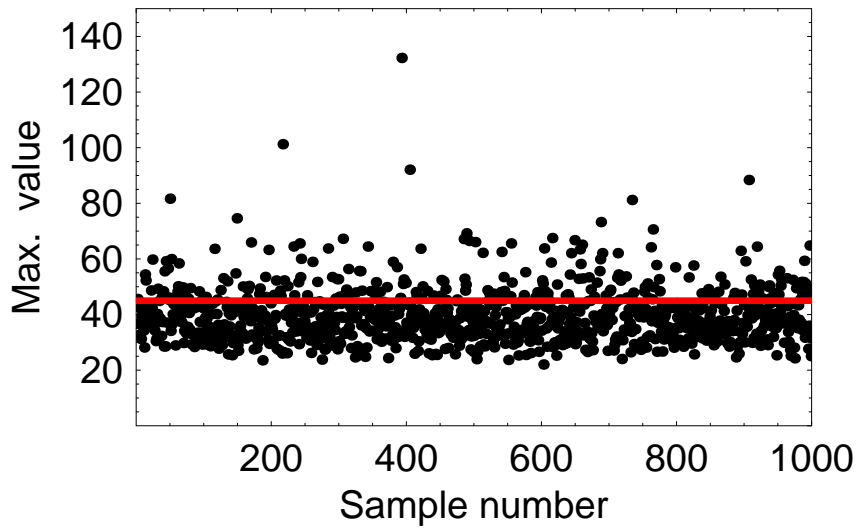}}\protect
\vskip 0.2cm \protect \caption{\label{fig1}{\footnotesize Maximal
values of $1000$ samples of size $N=59$. The maximal value of the
basic sample is $y_{0}(59) \approx 45$ and denoted by red line.}}
\end{figure}

\noindent The maximal values of samples are shown in Fig.
\ref{fig1}. The maximum in the basic sample is $y_{0}(59) \approx 45$.  The lowest of the maximal values in the experiment is $22.04$, while the
largest is $132.27$. One can observe that in $234$ cases
(more than $23${\%} of the one thousand samples) the maximum exceeds
 $y_{0}(59) \approx 45$.
\begin{figure} [ht!]
\protect \centering{
\includegraphics[scale=0.7]{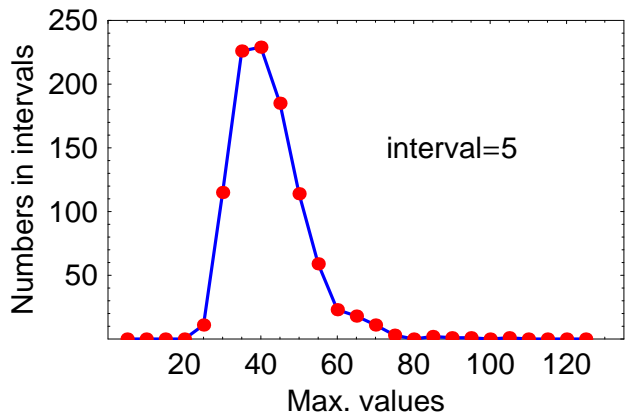}}\protect
\vskip 0.2cm \protect \caption{\label{fig1a}{\footnotesize
Distribution of the maximal values of $1000$ samples of size $N=59$
in intervals of length $5$.}}
\end{figure}
\begin{figure}[ht!]
\protect \centering{
\includegraphics[scale=0.7]{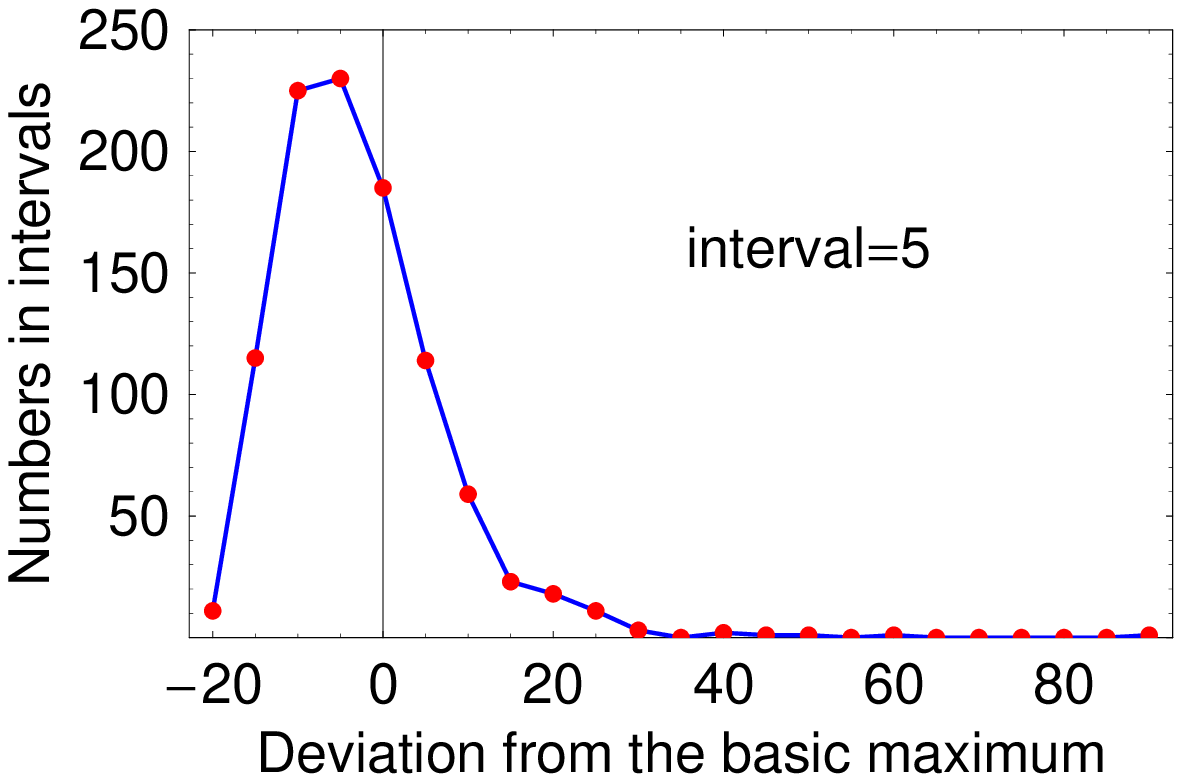}}\protect
\vskip 0.2cm \protect \caption{\label{fig1b}{\footnotesize
Distribution of the deviations of the maximal values from the basic
maximum $45$ in intervals of length $2.5$.}}
\end{figure}

In order to visualize better the random behavior of the maxima, we condensed the maxima, and counted the frequency of maxima falling into $[(i-1)*5,i*5], i=1,\dots 200$. That empirical distribution is shown in
Fig. \ref{fig1a}. Fig. \ref{fig1b} shows the difference $y_k(59)-y_0(59)$.

Let us check now if these properties are comply with the statistics. The
probability that the largest element in a given sample is greater
than $Q_{\gamma}$ is $1-\gamma^{N}$. Let $\xi _{n}(Q_{\gamma})$
stand for the random variable giving the number of maximum elements
exceeding $Q_{\gamma}$. The probability distribution of the newly
introduced random variable is
\begin{equation} \label{11}
{\mathcal P}\{\xi_{n}(Q_{\gamma}) = k\} = \binom{n}{k}
\;(1-\gamma^{N})^{k}\;\gamma^{N(n-k)}.
\end{equation}
From this expression we obtain the expectation value and the
variance as
\begin{equation} \label{12}
{\mathbf E}\{\xi_{n}(Q_{\gamma})\} = n(1-\gamma^{N}),
\end{equation}
\begin{equation} \label{13}
{\mathbf D}^{2}\{\xi_{n}(Q_{\gamma})\} =
n\;\gamma^{N}\;(1-\gamma^{N}).
\end{equation}
When $n$ and $k$ are sufficiently large, the distribution of the
random variable
\begin{equation} \label{14}
\chi_{n}(Q_{\gamma}) = \frac{\xi_{n}(Q_{\gamma}) - {\bf
E}\{\xi_{n}(Q_{\gamma})\}}{{\bf D}\{\xi_{n}(Q_{\gamma})\}}
\end{equation}
is approximately standard normal, hence,
\begin{equation} \label{15}
{\mathbf E}\{\xi_{n}(Q_{\gamma})\} - \kappa\;{\mathbf
D}\{\xi_{n}(Q_{\gamma})\} \leq \xi_{n}(Q_{\gamma}) \leq {\mathbf
E}\{\xi_{n}(Q_{\gamma})\} + \kappa\;{\mathbf
D}\{\xi_{n}(Q_{\gamma})\}
\end{equation}
is valid with probability $w$, and $\kappa $ is the root of the
equation
\begin{equation} \label{16}
\frac{1}{\sqrt{2\pi}}\;\int_{-\infty}^{\kappa} e^{-u^{2}/2}\;du =
\frac{1+w}{2}.
\end{equation}
Substituting here $n=1000$, $N=59$, $\gamma=0.95$ and $w=0.95$, we
get $\mathbf{E}\{\xi_{n}(Q_{\gamma})\}\approx 952, \;
\mathbf{D}\{\xi _{n}(Q_{\gamma})\} \approx 6.79$, and $\kappa
\approx 1.96$. The inequality (\ref{15}) is given by $$938 <
\xi_{1000}(Q_{0.95}) < 964,$$ where $\xi_{1000}(Q_{0.95}) \approx
940$. This relationship is witnessing the correctness of the
statistics. In spite of the nice agreement we wish to underline that
the $0.95 \vert 0.95$ methodology does not exclude rare events such
as limit violation when some of the calculated values are over the
technological limit $U_{T}$.

\begin{figure}[ht!]
\protect \centering{
\includegraphics[scale=0.7]{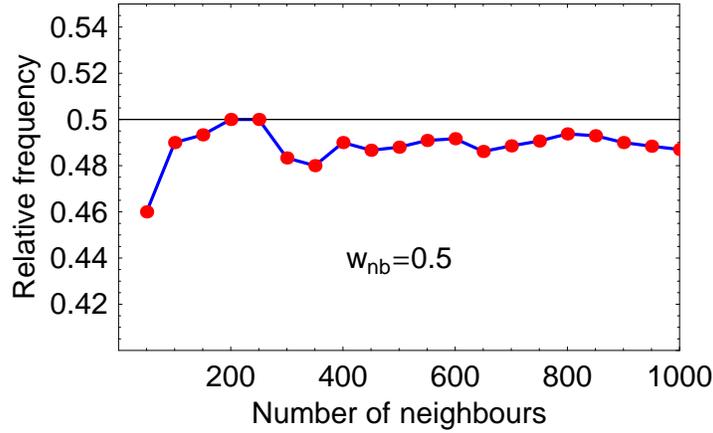}}\protect
\vskip 0.2cm \protect \caption{\label{fig1c}{\footnotesize Relative
frequency of the event $y_{\ell-1}(59) < y_{\ell}(59)$ in $50, 100,
\ldots, 1000$ neighbors.}}
\end{figure}

To illustrate that the probability of finding a larger maximal value
in a sample than in the previous one, is $w_{nb} = 0.5$, we have
used the maximums in $1000$ samples of size $59$ and have calculated
the relative frequencies of the event $y_{\ell-1}(59) <
y_{\ell}(59)$ in $50(50)1000$ neighbors. The results of calculations
are seen in Fig. \ref{fig1c}.

One should emphasize that this behavior of the upper limit of the
tolerance interval is not surprising, \textbf{this is a well established
consequence of the random nature of  the maximal values selected
from the samples}. It seems to be justified the need that the
tolerance interval method has to be replaced by a more reliable
method, \textbf{if it is possible at all}.

\section{Method based on sign test}

The concluding remarks at the end of the previous section are not optimistic. The question whether one can find a suitable check, which is based on a computer model, on the safety of a large device cannot be answered satisfactorily. Below we propose a method, borrowed from Kendall's book~\cite{kend79}, called "sign test" which may be more adequate than the 0.95/0.95 method.
\subsection{Single output variable}

Again, we assume the cumulative distribution function $G(y)$ of
the output variable to be continuous but unknown. Let
$S_{N}=\{y_{1},\ldots, y_{N}\}$ be a sample of $N$ observations
(runs of a computer model). Define the function
\begin{equation} \label{17}
\Delta(x) = \left\{ \begin{array}{ll}
1, & \mbox{if $x > 0$,} \\
\mbox{} & \mbox{} \\
0, & \mbox{if $x < 0$,} \end{array} \right.
\end{equation}
and introduce the statistical function
\begin{equation} \label{18}
z_{N} = \sum_{k=1}^{N} \Delta(U_{T} - y_{k})
\end{equation}
which gives the number of sample elements smaller than the safety
level $U_{T}$ by declaring that any state of the device with output
smaller than $U_{T}$ is safe. Criteria based on statistical function
(\ref{18}) are called sign criterion since $z_{N}$ counts only the
positive $U_{T} - y_{k}$ differences. When $G(y)$ is continuous, the
probability of $U_{T} - y = 0$ is surly zero.

Obviously, distribution of $z_{N}$ is binomial, using the notation
\begin{equation} \label{19}
{\mathcal P}\{\Delta(U_{T}-y) = 1\} = {\mathcal P}\{y
\leq U_{T}\} = p,
\end{equation}
we obtain
\begin{equation} \label{20}
{\mathcal P}\{ z_{N} = k\} = \binom{N}{k}\; p^{k}\;(1-p)^{N-k},
\;\;\;\;\;\;  k = 0, 1, \ldots, N.
\end{equation}
The decisive parameter is the probability $p$ which can be called
safety probability, and our task is to find a confidence interval
$[\gamma_{L}(k),\;\gamma _{U}(k)]$ that covers the value $p$ with a
prescribed probability $\beta$ provided we have a sample of size $N$
and in the sample $z_{N}=k \leq N$ elements smaller than the safety
level $U_{T}$. The mathematics of the problem is well known, and one
can find a large number of publications \cite{wils27} -
\cite{newc98} in this field. However, there are several special
aspects of the safety analysis, which need some further
considerations.

Clearly, the formula (\ref{19}) gives the probability that the
output $y$ is not larger than the safety level $U_{T}$. When the
lower limit $\gamma_{L}(k)$ of the confidence interval is close to
unity, we can claim at least with probability $\beta$ that the
chance of finding the output $y$ smaller than $U_{T}$ is also close
to unity and the device under consideration, in its output $y$, can
be regarded as safe at the level $[\beta \vert \gamma_{L}(k)]$.

If the sample size $N > 50$, the random variable
\begin{equation} \label{21}
\frac{k - Np}{\sqrt{Np\;(1 - p)}} = \zeta_{k}
\end{equation}
has approximately normal distribution. Here $k$ is the number of
sample elements not exceeding $U_{T}$. Let $\beta$ denote the
confidence level, then
\[ {\mathcal P}\{\vert \zeta_{k} \vert \leq u_{\beta}\} =
2\Phi(u_{\beta}) - 1 = \beta, \] where $\Phi(x)$ is the standard
normal distribution function. This equation can be rewritten in the
form
\[ {\mathcal P}\{\vert \zeta_{k} \vert \leq u_{\beta}\} =
{\mathcal P}\{ (N+u_{\beta}^{2})(p-\gamma_{L})(p-\gamma_{U}) \leq
0\} = \beta,\] where
\[ \gamma_{L} = \gamma_{L}(k, u_{\beta}) = \]
\begin{equation}\label{22}
= \frac{1}{N+u_{\beta}^{2}}\;\left[k + \frac{1}{2}u_{\beta}^{2} -
u_{\beta}\sqrt{k(1-k/N) + u_{\beta}^{2}/4}\right],
\end{equation}
and
\[ \gamma_{U} = \gamma_{U}(k, u_{\beta}) = \]
\begin{equation}\label{23}
= \frac{1}{N+u_{\beta}^{2}}\;\left[k + \frac{1}{2}u_{\beta}^{2} +
u_{\beta}\sqrt{k(1-k/N) + u_{\beta}^{2}/4}\right].
\end{equation}
Here $u_{\beta}$ is the root of
\[ \Phi(u_{\beta}) = \frac{1}{2}(1 + \beta). \]

In a number of cases it suffices to know the probability of the
event $\{\gamma _{L}(k, v_{\beta}) \leq p\}$, where $v_{\beta} =
\Phi^{-1}(\beta)$.  Since $\zeta_{k}$ with fixed  $k$ is a
decreasing function of $p$, the events $\{\zeta_{k } \leq
v_{\beta}\}$ and $\{\gamma _{L}(k, v_{\beta}) \leq p\}$ are
equivalent, hence \[ {\mathcal P}\{ \zeta_{k} \leq v_{\beta}\} =
{\mathcal P}\{ \gamma_{L}(k, v_{\beta}) \leq p\} = \Phi(v_{\beta}) =
\beta.\] Consequently, the operation of a device can be regarded
safe if the parameter $p$ for all output variables is covered by
$[\gamma_{L}(k, v_{\beta }), 1]$ with a prescribed probability
$\beta$, provided that $\gamma_{L}(k, v_{\beta})$ is close to unity.

\begin{table}[ht!]
\caption{\label{tab1} {\footnotesize Number of successes $k$ in a
sample of size $N$}} \vspace{0.2cm}
\begin{center}
\begin{tabular}{|c|c|c|c|c|c|c|c|c|c|c|c|c|} \hline
$k$ & 99 & 108 & 118 & 128 & 137 & 147 & 157 & 166 & 176 & 185 & 195
\\ \hline
$N$ & 100 & 110 & 120 & 130 & 140 & 150 & 160 & 170 & 180 & 190 &
200  \\ \hline
\end{tabular}
\end{center}
\end{table}
\noindent Table \ref{tab1} gives the number of successes $k$ in a
sample of size $N$ needed for acceptance at the level $\beta =
\gamma_{L} = 0.95$. We utilized the approximate formula (\ref{22})
with $u_{\beta} \rightarrow v_{0.95} \approx 1.64$ to derive the
entries in Table \ref{tab1}.

\begin{figure} [ht!]
\protect \centering{
\includegraphics[scale=0.9]{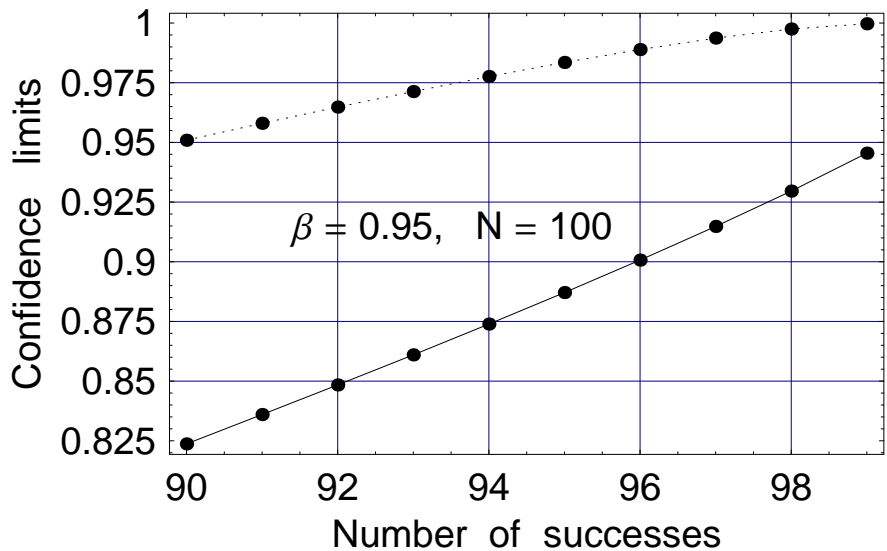}}\protect
\vskip 0.2cm \protect \caption{\label{fig2}{\footnotesize
Dependence of $\gamma_{L}$ and $\gamma_{U}$ on the number of
successes in a sample of N=100 elements.}}
\end{figure}

When the sample size is less than $50$, we may not apply the
asymptotically valid normal distribution. The below given derivation
of the confidence limits is a modified version of the method
proposed by Clopper and Pearson \cite{clpr34}. The probability of at
least $k$ successes from $N$ observations is given by
\[ S_{k}^{(N)}(p) = \sum_{j=0}^{k} \binom{N}{j}\;
p^{j}\;(1-p)^{N-j},\] where $p = {\mathcal P}\{y \leq U_{T}\}$.
This formula can be recast as
\[ S_{k}^{(N)}(p) =
\frac{N!}{k!\;(N-k-1)!}\;\int_{0}^{1-p} (1-v)^{k}\;v^{N-k-1}\;dv,\]
and it is clear from that expression that $S_{k}^{(N)}(p)$ is a
monotonously decreasing function of $p$. Since
\[ S_{k}^{(N)}(p) = \left\{ \begin{array}{ll}
1, & \mbox{if $p=0$,} \\
\mbox{} & \mbox{} \\
0, & \mbox{if $p=1$,} \end{array} \right. \]  it assumes an
arbitrary value only once in the interval [0,1]. Consequently, a
$p = p_{\delta}$ value exists so that
\[ S_{k}^{(N)}(p_{\delta}) = \delta, \;\;\;\;\;\; \forall \; 0
< \delta < 1. \] Exploiting the monotony, we can construct a
function such that
\[  R_{k}^{(N)}(p) < R_{k}^{(N)}(p_{\delta}) = \delta,\]
when $p > p_{\delta}$. Such a function is
\[ R_{k}^{(N)}(p) = 1 - S_{k-1}^{(N)}(p) = \sum_{j=k}^{N}
\binom{N}{j}\; p^{j}\;(1-p)^{N-j}, \] Finally, we establish the
upper limit $\gamma_{U}$ from
\begin{equation} \label{23a}
S_{k}^{(N)}(\gamma_{U}) \leq \frac{1}{2}(1-\beta),
\end{equation}
and the lower limit $\gamma _{L}$ from
\begin{equation} \label{23b}
R_{k}^{(N)}(\gamma_{L}) \leq \frac{1}{2}(1-\beta).
\end{equation}
The interval $[\gamma_{L}, \; \gamma_{U}]$  covers the unknown
parameter $p$ with probability $\beta $. The dependence of $\gamma
_{L}$ and $\gamma _{U}$ are shown in Fig. \ref{fig2} for a sample of
$N=100$ elements, $\beta$ stands for confidence level.

\subsection{Several Output Variables}

Now we assume the output to comprise $\ell$ variables. Let these
variables be $y_{1}, \ldots, y_{\ell}$. There are several fairly
good test to prove if they are statistically independent or not. To
each of the independent variables we can apply the considerations
above but for dependent variables we need novel approach. Let
\[ \mathbf{S}_{N} = \left(\begin{array}{cccc}
y_{1 1} & y_{1 2} & \ldots & y_{1 N} \\
y_{2 1} & y_{2 2} & \ldots & y_{2 N} \\
\vdots & \vdots & \ddots & \vdots \\
y_{\ell 1} & y_{\ell 2} & \ldots & y_{\ell N}
\end{array} \right) \]
denote the sample matrix obtained in $N >> \ell$ independent
observations. With a computer model, an observation is a run.
Introducing the column vector $\vec{y}_{k}$ of $\ell$ components,
the sample matrix can be written as
\[ \mathbf{S}_{N} = \left(\vec{y}_{1}, \ldots, \vec{
y}_{N}\right). \] In accordance with our assumption, different
vectors $\vec{y}_{k}$ are statistically independent but the
components in a given vector may be dependent.

Below we expound the sign test only for two output variables $y_{1}$
and $y_{2}$ relying on the assumption that their joint distribution
function $G(y_{1}$,$y_{2})$ is unknown but continuous in either
variable. The goal of the foregoing analysis is to verify the safety
conditions  $y_{1} < U_{T}^{(1)}$ and $y_{2} < U_{T}^{(2)}$. When
the conditions are accomplished with probability $p_{12} =
G(U_{T}^{(1)}, U_{T}^{(2)}) \approx 1$ we say the device is safe for
the outputs $y_{1}$ and $y_{2}$. Here, as before, the limits
$U_{T}^{(1)}$, and $U_{T}^{(2)}$ are determined by technological
requirements. Since $p_{12}$ is unknown, our job is to construct a
confidence interval $[\gamma_{L}^{(1,2)},\; \gamma_{U}^{(1,2)}]$ so
that it covers $p_{12}$ with probability $\beta_{12}$. In most cases
it suffices to calculate solely the lower confidence limit
$\gamma_{L}^{(1,2)}$, and to use the interval
$[\gamma_{L}^{(1,2)},\;1]$ as confidence interval.

The event $\{y_{1k} < U_{T}^{(1)},\; y_{2k} < U_{T}^{(2)}\}$ will be
called a success. Now, introduce the statistical function
$$ z_{N}^{(1,2)} = \sum_{k=1}^{N}
\Delta(U_{T}^{(1)}-y_{1k})\;\Delta(U_{T}^{(2)}-y_{2k}) $$ which
gives the number of successes in a sample of size $N$. Obviously, if
$y_{1k} < U_{T}^{(1)}$ and $y_{2k} < U_{T}^{(2)}$, then
$$ \Delta(U_{T}^{(1)}-y_{1k})\;\Delta(U_{T}^{(2)}-y_{2k}) = 1,$$
while $0$ otherwise. Since the newly introduced random variable
$z_{N}^{(1,2)}$ is the sum of $N$ independent random variables,
assuming values either 1 or 0, its distribution is binomial. Using
the notation
\[ {\mathcal P}\{\Delta(U_{T}^{(1)}-y_{1})\;
\Delta(U_{T}^{(2)}-y_{2})=1\} = {\mathcal P}\{y_{1} < U_{T}^{(1)},
\; y_{2} < U_{T}^{(2)}\} = p_{12},  \] we can write
\[ {\mathcal P} \{z_{N}^{(1,2)}=k \} = \binom{N}{k}\;
p_{12}^{k}\;(1-p_{12})^{N-k},\] for $k = 0, 1, \ldots, N$. Clearly,
$p_{12}$ is the joint probability of safety for the output variables
$y_{1}$ and $y_{2}$.

At this point we rejoin the thought of line of the previous
subsection. Instead of repeating the already familiar argumentation,
we amend two trivial although important remarks. Let us define the
following two statistical functions:
$$ z_{N}^{(1)} = \sum_{i=1}^{N}
\Delta(U_{T}^{(1)} - y_{1i}) \;\;\;\;\;\; \mbox{and} \;\;\;\;\;\;
z_{N}^{(2)} = \sum_{j=1}^{N} \Delta(U_{T}^{(2)} - y_{2j}). $$ In
general, these two functions are statistically not independent. Each
of them is a sum of $N$ independent random variables with values 1
or 0, therefore, one can write $$ {\mathcal P}\{z_{N}^{(1)}=i\} =
\binom{N}{i}\; p_{1}^{i} (1-p_{1})^{N-i}$$ and $$ {\mathcal
P}\{z_{N}^{(2)}=j\} = \binom{N}{j}\; p_{2}^{j} (1-p_{2})^{N-j}, $$
$$ i,j = 1, \ldots, N, $$ where
\[ p_{\ell} = {\mathcal P}\{y_{\ell} < U_{T}^{(\ell)}\} =
{\mathcal P}\{\Delta(U_{T}^{(\ell)} - y_{\ell}) = 1\}, \;\;\;\;\;\;
 \ell = 1, 2, \]
are unknown probabilities of safety for output variables $p_{1}$ and
$p_{2}$. Applying the method used previously, this time separately
to the samples  \[ {\mathcal S}_{N}^{(1)} = \{y_{1i}, \;\; i=1,
\ldots, N\} \;\;\;\;\;\; \mbox{and} \;\;\;\;\;\; {\mathcal
S}_{N}^{(2)} = \{y_{2j}, \;\; j=1, \ldots, N\},\] we construct two
random intervals $[\gamma_{L}^{(1)}, \;1]$ and $[\gamma_{L}^{(2)},
\;1]$ covering $p_{1}$ and $p_{2}$ with probabilities $\beta_{1}$
and $\beta_{2}$, respectively. Evidently, it could occur that the
levels $(\beta_{1}\vert \gamma_{L}^{(1)})$ and $(\beta_{2}\vert
\gamma_{L}^{(2)})$ corroborate the claim that samples ${\mathcal
S}_{N}^{(1)}$ and ${\mathcal S}_{N}^{(2)}$ separately comply with
safety requirements. However, this does not mean that we would
arrive at the same conclusion from analyzing the two samples
jointly. The reason is that $y_{1}$ and $y_{2}$, the two output
random variables are statistically in general not independent.
Hence, we should ascertain weather the interval
$[\gamma_{L}^{(1,2)}, \;1]$ covers the probability $p_{12}$ with the
pre-assigned probability $\beta_{12}$.

Decision on the safety, when two output variables are subjected to
limitations should go as follows. Firstly, we test the hypothesis
concerning dependence of the output variables $y_{1}$ and $y_{2}$.
If they are dependent, we should estimate the random interval
$[\gamma_{L}^{(1,2)}, \, 1]$ which covers $p_{12}$ with probability
$\beta_{12}$. Solely if they are statistically independent should we
estimate the random intervals $[\gamma_{L}^{(1)},\, 1]$ and
$[\gamma_{L}^{(2)},\, 1]$  covering $p_{1}$ and $p_{2}$ with
probabilities $\beta_{1}$ and $\beta_{2}$, respectively.

For the sake of demonstration, an example is given below. First, the
lower confidence limits are calculated for the parameter $p$ as a
function of the success number $k$, in the case of sample size
$N=100$, at the confidence levels $\beta = 0.9, 0.95$ and $0.99$.

\begin{center}
\begin{table}[ht!]
\caption{\label{tab2}{\footnotesize Lower confidence limits in a
sample of $N=100$ as a function of the success number $k$ at three
confidence levels $\beta$}} \vspace{0.3cm}
\begin{center}
\vspace{0.3cm}
\begin{tabular}{|c|c|c|c|} \hline
$k$ $\backslash$ $\beta$ & 0.90 & 0.95 & 0.99
\\ \hline 90 & 0.8501 & 0.8362 & 0.8086
\\ \hline 91 & 0.8616 & 0.8482 & 0.8212
\\ \hline 92 & 0.8733 & 0.8602 & 0.8340
\\ \hline 93 & 0.8850 & 0.8725 & 0.8471
\\ \hline 94 & 0.8970 & 0.8850 & 0.8604
\\ \hline 95 & 0.9092 & 0.8977 & 0.8741
\\ \hline 96 & 0.9216 & 0.9108 & 0.8882
\\ \hline 97 & 0.9344 & 0.9242 & 0.9030
\\ \hline 98 & 0.9476 & 0.9383 & 0.9185
\\ \hline 99 & 0.9616 & 0.9534 & 0.9354
\\ \hline 100 & 0.9772 & 0.9704 & 0.9549 \\ \hline
\end{tabular}
\end{center}
\end{table}
\end{center}

Then, by using Monte Carlo simulation, we have generated two samples
a) and b) either sample contains $N=100$ observations (runs) of two
output variables. The samples have been generated from a bivariate
normal distribution with parameters $m_{1}=m_{2}=0$ and
$\sigma_{1}=\sigma_{2}=1$ but the correlation coefficient is $C=0.1$
in the sample a) while that is $C=0.7$ in the sample b). The
acceptance range is $[-2,2]$ for both output variables. In sample a)
and b) $4$ and $2$ samples lie respectively outside the acceptance
range.

\begin{figure} [ht!]
\protect \centering{
\includegraphics[scale=0.9]{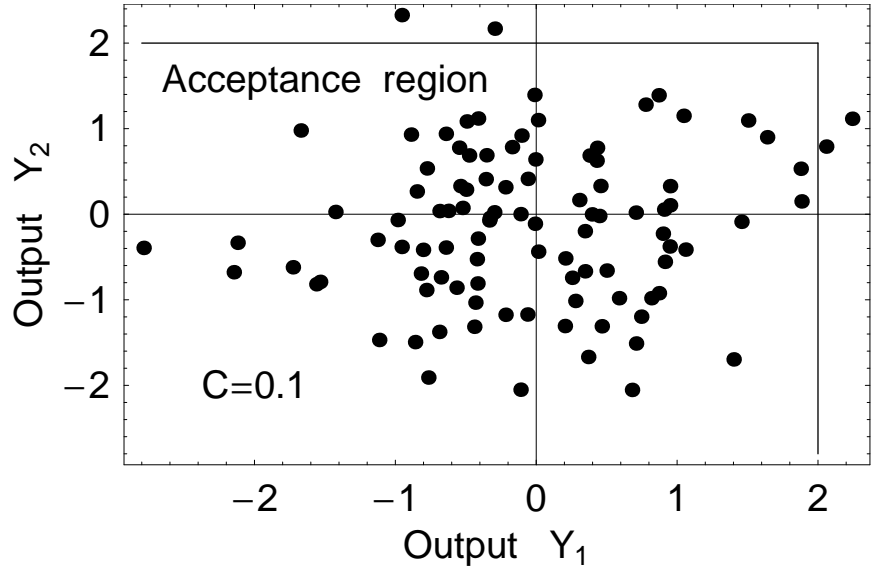}}\protect
\vskip 0.2cm \protect \caption{\label{fig3}{\footnotesize Sample
a)}}
\end{figure}
\vspace{-0.1cm}
\begin{figure} [ht!]
\protect \centering{
\includegraphics[scale=0.9]{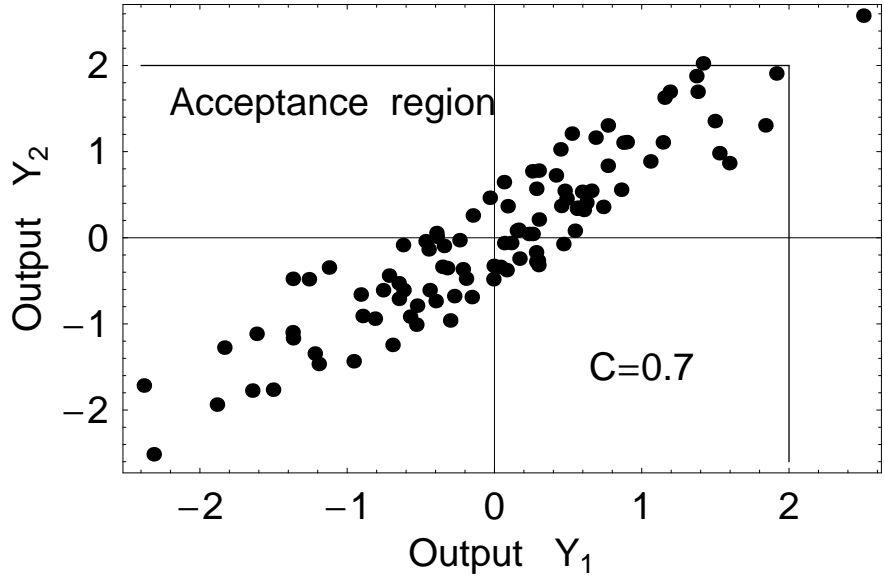}}\protect
\vskip 0.2cm \protect \vspace*{-0.2cm}
\caption{\label{fig4}{\footnotesize Sample b)}}
\end{figure}

First let us consider the sample a). It can be seen that the number
of successes among $100$ events is $96$, i.e. the number of safety
violation is $4$. From Table \ref{tab2} one can read in this case
that the interval [0.9108,1] covers the parameter $p_{12}$ with
probability $\beta_{12}=0.95$. This coverage is rather small to
state the device is safe. When we assess the output variables one by
one, we see that the associated parameters $p_{1}$ and $p_{2}$ are
covered by the interval $[0.9383,1]$ with probability $\beta=0.95$
in either sample. However tempting is to accept $0.9383$ as lower
bound for the probability to be used in safety analysis, that number
has nothing to do with $p_{12}$ and should not be used in safety
analysis.

Now let us pass on to sample b) where we see a strong correlation
between $y_{1}$ and $y_{2}$. The number of safety violations is $2$
and from Table \ref{tab2} one can read that the confidence interval
$[0.9383,1]$ covers the parameter $p_{12}$ with probability
$\beta_{12}=0.95$. In other words, we conclude that the probability
of the event $\{y_{1} < U_{T}^{(1)},\; y_{2} < U_{T}^{(2)}\}$ is at
least $0.9383$. Since there is only $1$ violation for both output
variables $y_{1}$ and $y_{2}$, according to Table \ref{tab2} the
parameters $p_{1}$ and $p_{2}$ are covered by the same interval
$[0.9534,1]$ on the level $\beta_{1}=\beta_{2}=0.95$. Again, however
favorable these numbers are, they should not be used in assessing
safety. The above discussed simple numerical example clearly
indicated the danger awaiting the analyst when his/her judgment is
based on tests performed separately on correlated output variables.

Finally, we mention that the generalization of the sign test to
$\ell>2$ output variables is straightforward, we have to use the
statistical function
\begin{equation} \label{24}
z_{N}^{(1,\ldots,\ell)} = \sum_{k=1}^{N}\;\prod_{j=1}^{\ell}
\Delta(U_{T}^{(j)}-y_{jk})
\end{equation}
to evaluate safety based on observation of $N$ samples of the $\ell$
output variables. In this manner we obtain the sum of $N$
independent random variables in expression (\ref{24}), and then, the
further steps will be the same as at the beginning of the
subsection.

\subsection{Criticism of the sign test method}

It is evident that the sign test method is based on the interval
estimation of the binomial proportion. Therefore, all difficulties
connected with the erratic behavior of the coverage probability of
the confidence interval are appearing in the sign test method, too.

Let $N$ be the number of runs and $k$ is the number of events $\{y <
U_{T}\}$, i.e. the number of successes. Denote by ${\mathcal
D}_{\alpha}(N, k)$ the set of points of the confidence interval
$$[\gamma_{L}(N, k, \alpha), \,\gamma_{U}(N, k, \alpha)],$$ where
$\alpha = 1 - \beta$ and $\beta$ is the nominal coverage
probability, i.e. the coincidence level. Introduce the indicator
function
\begin{equation} \label{25}
\mathbb{I}_{\alpha}(N, k, p) = \left\{ \begin{array} {ll} 1, &
\mbox{if $p
\in {\mathcal D}_{\alpha}(N, k)$,} \\
\mbox{ } & \mbox{ } \\
0, & \mbox{if $p \not\in {\mathcal D}_{\alpha}(N, k)$}
\end{array} \right.
\end{equation}
and define the coverage probability by the relationship
\begin{equation} \label{26}
C_{\alpha}(N, p) = \sum_{k=0}^{N}\mathbb{I}_{\alpha}(N, k,
p)\,{\mathcal P}\left\{z_{N}=k\right\} =
\sum_{k=0}^{N}\mathbb{I}_{\alpha}(N, k,
p)\,\binom{N}{k}\,p^{k}\,(1-p)^{N-k},
\end{equation}
which, as we will see, is different from the nominal coverage
probability $\beta$.

In order to  characterize the quality of the coverage it is often
used the mean coverage probability
\begin{equation} \label{27}
\langle C_{\alpha}(N, p) \rangle = \int_{0}^{1} C_{\alpha}(N,
p)\;dp,
\end{equation}
and the mean length of the confidence interval
\begin{equation} \label{28}
\mathbb{L}_{\alpha} = \sum_{k=0}^{N} \left[\gamma_{U}(N, k, \alpha)
- \gamma_{L}(N, k, \alpha)\right]\,\binom{N}{k}\,p^{k}\,(1-p)^{N-k}.
\end{equation}
In the case of the Clopper-Pearson interval, which we introduced by
(\ref{23a}) and (\ref{23b}), the actual coverage probability is
always equal to or above the confidence level $\beta$.

In the practice, it is often used the one-sided confidence interval
$\left\{\gamma_{L}(N, k, \alpha), \, 1\right\}$, since the usual
question is wether the $\gamma_{L}(N, k, \alpha)$ value is large
enough for the safety parameter $p$ at a prescribed confidence level
$1 - \alpha = \beta$. It can be shown that the root $p=p_{L}$ of the
equation
\begin{equation} \label{29}
I_{p}(k, N-k+1) = \alpha
\end{equation}
is the lower confidence limit, i.e.  $p_{L} = \gamma_{L}(N, k,
\alpha)$. Here, $I_{p}(k, N-k+1)$ is the cumulative
beta-distribution function.
\begin{table}[ht!]
\caption{\label{tab3} {\footnotesize Dependence of the lower
confidence limit $\gamma_{L}(N, k, \alpha)$ on the number of
successes $k$ at the sample size $N=100$} and the significance level
$\alpha=0.05$} \vspace{0.1cm}
\begin{center} \footnotesize{
\begin{tabular}{|c|c|c|c|c|c|c|c|c|c|c|c|c|} \hline
$k$ & 90 & 91 & 92 & 93 & 94 & 95 & 96 & 97 & 98 & 99 & 100
\\ \hline
$\gamma_{L}$ & 0.8363 & 0.8482 & 0.8603 & 0.8725 & 0.8850 & 0.8977 &
0.9108 & 0.9243 & 0.9384 & 0.9534 & 0.9705  \\ \hline
\end{tabular}}
\end{center}
\end{table}
From Table \ref{tab3} one can see that at least $99$ successes out
of $100$ are needed to state the safety parameter $p$ is higher than
$0.95$ with probability $0.95$. In order to demonstrate the erratic
behavior of the coverage probability of the confidence interval
$\left\{\gamma_{L}(N, k, \alpha), \, 1\right\}$, we calculated the
sum $$C_{0.05}(100, p) = \sum_{k=0}^{100}\mathbb{I}_{0.05}(100, k,
p)\,\binom{100}{k}\,p^{k}\,(1-p)^{100-k},$$ where now
$\mathbb{I}_{0.05}(100, k, p)$ is defined on the points of the
interval $\left\{\gamma_{L}(100, k, 0.05), \, 1\right\}$.

\begin{figure} [ht!]
\protect \centering{
\includegraphics[scale=0.9]{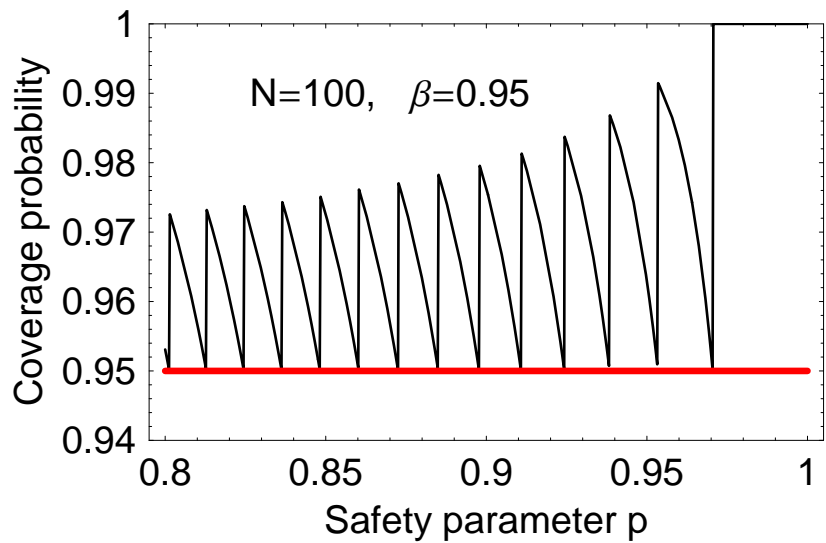}}\protect
\vskip 0.2cm \protect \caption{\label{fig5}{\footnotesize Dependence
of the coverage probability on the safety parameter at the sample
size $N=100$ and at the nominal confidence level $\beta=0.95$)}}
\end{figure}

\begin{figure} [b]
\protect \centering{
\includegraphics[scale=0.7]{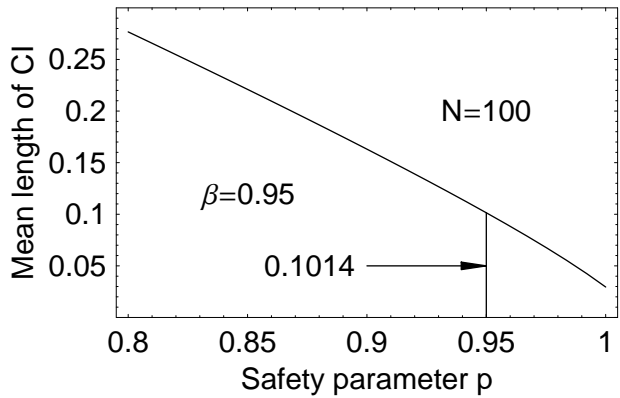}}\protect
\vskip 0.2cm \protect \caption{\label{fig6}{\footnotesize Dependence
of the mean length of CI on the safety parameter at the sample size
$N=100$ and at the nominal confidence level $\beta=0.95$)}}
\end{figure}

\noindent In Fig. \ref{fig5} one can see that for any fixed $p$ the
actual coverage probability can be much larger than $\beta = 1 -
\alpha$. It means that the Clopper-Pearson interval is very
conservative and is not a good choice for practical use, unless the
prescription $C_{\alpha}(n, p) \geq \beta$ is demanded. As known,
there are many other intervals \cite{brown01}  for the estimation of
$p$, however, the erratic behavior of the coverage probability
cannot be ceased.

By using the formula (\ref{28}) one can plot the dependence of the
mean length of confidence interval on the safety parameter $p$. In
Fig. \ref{fig6} it can be seen this dependence in the case of the
sample size $N=100$ and the confidence level $\beta=0.95$,
respectively. It is interesting to note, that the mean lower
confidence limit $\langle \gamma_{L}(100, k, 0.05)\rangle \approx 1
- 0.1014 = 0.8986$ corresponds nearly to the success number $k=95$.

\section{Simple considerations}

The safety analysis of an industrial device consists of the
following steps. The designer determines the nominal state of the
device, and enlists the parameters influencing the state of the
device. We call those parameters input. Either the designer or the
analyst determines the probability distribution of the input,
usually by engineering judgment. The analyst selects a code to carry
out simulation of the device operation, and runs the code with a
given number of random (or other reasonably chosen) inputs.
Therefore, the output variables should have a random part, and the
analyst tries to evaluate these uncertainties, by using for instance
the tolerance interval method. In order to illustrate some problems
of the statistical approach to be applied in safety analysis, below
we define an over simplified device.

\begin{figure}[ht!]
\centering{
\includegraphics[scale=0.7]{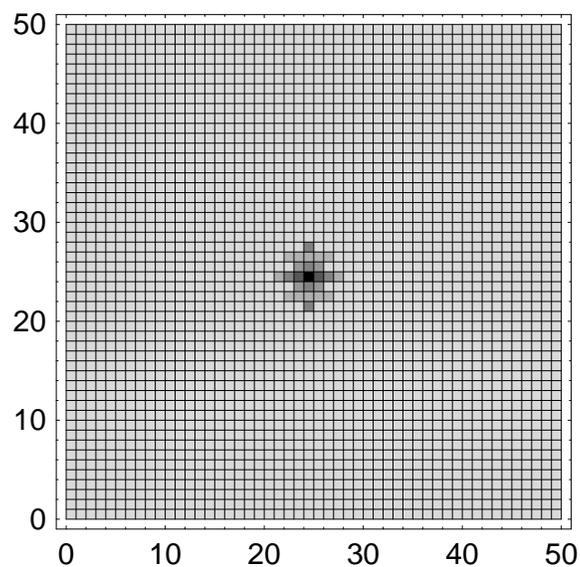}}
\caption{\label{fig8} \footnotesize{Risks associated with cells}}
\end{figure}

Assume that the device is represented by $50 \times 50 = 2500$ cells
each of which corresponds to a possible state of the device. Each
state is determined by a fixed real number $0 \leq R \leq 1$ which
is identified as a scaled risk of the state. According to this,
every cell is characterized by a risk value. The operation of the
device consists of a random selection of one cell the $R$ value of
which defines the output variable $y$. The run of the "code"
corresponding to the best estimate method in this gedanken
experiment means the realization of this operation. Repeating the
run $N$-times, we obtain a sample with $N$ elements.\footnote{It is
important to note that this sample belongs to the family of discrete
distribution samples. However, it can be shown that the Wilk's
formula can be applied also in this case provided that the number of
discontinuities of the unknown cumulative distribution function is
finite.}

If $R=0$, then the state is safe, there is no limit violation. We
suggest to investigate a device where the limit violation may occur
only in a small fraction of the states, actually in 1\% of the
states. To simplify the presentation, the $R$ values are  given in
Figure \ref{fig8}.

\begin{figure}[ht!]
\centering{
\includegraphics[scale=0.7]{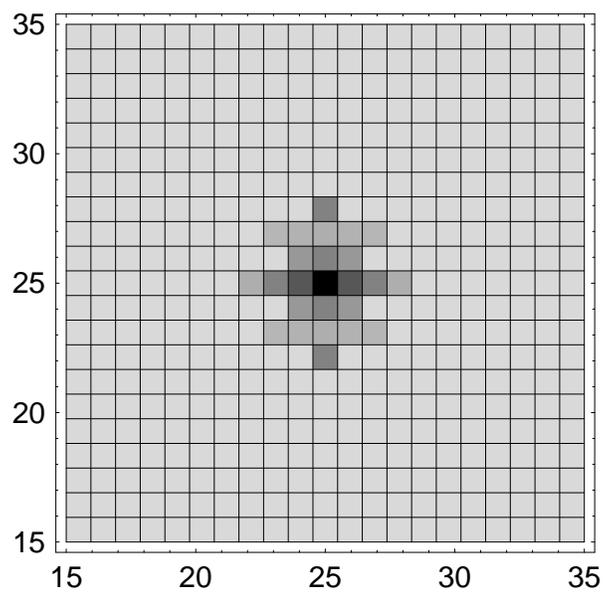}}
\caption{\label{fig9} \footnotesize{Central part of Fig.
\ref{fig8}.}}
\end{figure}
Most of the cells has no limit violation (light cells). Limit
violation is observed only at a few cells (25 of them, 1 \% of the
states): 12 cells have minor limit violations ($R \leq 0.1$), their
partition: $R=0.08$ in 4 cells, $R=0.09$ in 4 cells, $R=0.1$ in 4
cells. Further 10 cells have moderate limit violation ($0.1 \leq R
\leq  0.2$). Their partition is $R= 0.15$ in 4 cells, $R= 0.2$ in 6
cells, and 3 have severe limit violations ($ > 0.25$), their
partition is $R= 0.3$ in 2 cells, and $R= 0.5$ in 1 cell. All the
cells with limit violation are in the center of the region shown in
Fig. \ref{fig8}, see the enlargement in Fig. \ref{fig9}.

The $0.95 \vert 0.95$ methodology is performed in the following way:
one generates $N=59$ random integers from the set of uniformly
distributed integers $1, 2, \ldots, 2500$, and then reads the risk
values $R$ of the generated $59$ cells.~\footnote{During the
calculations the random number generator should be reseeded.} These
risk values form the basic sample. The maximal element $R_{1}(59)$
of the sample defines the random interval $[0,\, R_{1}(59)]$ which
covers the $95$ \% of the risk distribution with probability
$\beta=0.95$. Let us repeat this procedure $n=100$ times, and
determine the numbers of occurrences of the risk values $0, 0.08,
0.09, 0.1, 0.15, 0.2, 0.3, 0.5$. In order to see the effect of the
strengthening requirements, we performed the same calculations in
the cases of $N=90$ and $N=459$, which sample sizes correspond to
the levels $0.95 \vert 0.99$ and the $0.99 \vert 0.99$,
respectively.

\begin{table}[ht!]
\caption{\label{tab4} Maximum values of $R$ in $n=100$ calculations
with sample sizes $N=59, 90, 458$ in each calculation.}
\vspace{0.2cm}
\begin{center}
\begin{tabular}{|c|r|r|r|} \hline
\mbox{} & \multicolumn{3}{|c|}{Number in 100 cases with } \vline  \\
\hline
$R$ & $N=59$ & $N=90$ & $N=458$   \\
\hline
0.50 & 5 & 2 & 27 \\
0.30 & 2 & 7 & 17 \\
0.20 & 15 & 16 & 33 \\
0.15 & 4 & 13 & 12 \\
0.10 & 7 & 4 & 7 \\
0.09 & 4 & 7 & 2 \\
0.08 & 8 & 6 & 1 \\
0.00 & 55 & 45 & 1 \\
\hline
\end{tabular}
\end{center}
\end{table}

Actual results are given in Table \ref{tab4}. If $N=59$, then in
55\% of the cases, we conclude (falsely) that the system is safe,
and only in 7\% of the samples we get an alarming large, $R>0.2$
value, from which the correct maximum is obtained in 5\% of the
cases. If $N=90$, then we see some moderate improvement: solely 45\%
of the cases show no limit violation and 9\% indicate alarming limit
violation. With $N=458$, only 1\% of the cases show the system to be
safe and 44\% of the cases indicate alarmingly high limit violation.

We admit that the presented gedanken experiment is a real challenge
since only one state out of 100 has appreciable limit violation, but
perhaps in a real safety analysis we encounter possible device
states which are safe in general except for a small fraction of the
possible states. Safety analysis is done to reveal such situations.
The above given considerations may give us an impression what the
"high probability" might be in 10CFR §50.46.
\par
The reader may think that this is an exceptional case. However, it
is fairly common that only a small fraction of the reasonable inputs
may lead to safety hazard. Another problem is, that if we have a
large number of unsafe cells, in the average, one out of 20 will
remain uncovered, this is the meaning of the 95\% confidence level.

\section{Concluding remarks}

In the validation and verification process of a code one carries out
a series of computations. The input data are not precisely
determined because measured data have an error, calculated data are
often obtained from a more or less accurate model. Some users of
large codes are content with comparing the nominal output obtained
from the nominal input, whereas \textbf{all the possible inputs should be
taken into account when judging safety}. At the same time, any
statement concerning safety must be aleatory, and its merit can be
judged only when the probability is known with which the statement
is true.

There are several statistical tools applicable in safety analysis.
Before choosing the appropriate statistical method, one has to look
at the physical model. If we assume a situation where most of the
possible states are safe and our task is to pinpoint the low number
of possible dangerous states, \textbf{we need a larger statistical sample},
i.e. we have to carry out a larger number of calculations with the
computer model.

On the other hand, if there are many risky states, the tolerance
interval may not reveal all of them, there is a good chance that
every 20th risky cell remains undetected when the tolerance level is
0.95.

The authors arrived at conclusion that \textbf{the random character of the
computed output values is usually neglected by the analysts}. For
instance, the trivial fact that the maximal value of an output
variable obtained by the tolerance interval method is not
repeatable, since it is a random variable, is almost forgotten in
the applied uncertainty analysis. It may happen that the authority
will carry out an independent safety analysis which may lead to a
smaller safety reserve in 50\% of the cases and the analyst may find
himself/herself in an inconvenient situation. When the output
variables are correlated, then the separate analysis of the
variables may result incorrect statements.

Consequent application of order statistics or the application of the
sign test may offer a way out of the present situation. The authors
are also convinced that efforts should be made
\begin{itemize}
\item to study the statistics of the output variables,
\item to study the occurrence of large fluctuations in the analyzed cases
\item to avoid  random quantities  acquiring essential role in safety analysis.
\end{itemize}

All these observations should influence, in safety analysis, the
application of best estimate methods, and underline the opinion that
any realistic modeling and simulation of complex systems must
include the probabilistic features of the system and the
environment.

\end{document}